\input harvmac
\input amssym
\input epsf

\def\unit{\relax{\rm 1\kern-.26em I}}
\def\nada{\relax{\rm 0\kern-.30em l}}
\def\tilde{\widetilde}



\def\IL{\relax{\rm I\kern-.18em L}}
\def\IH{\relax{\rm I\kern-.18em H}}
\def\IR{\relax{\rm I\kern-.18em R}}
\def\IC{\relax\hbox{$\inbar\kern-.3em{\rm C}$}}
\def\IZ{\relax\ifmmode\mathchoice
{\hbox{\cmss Z\kern-.4em Z}}{\hbox{\cmss Z\kern-.4em Z}}
{\lower.9pt\hbox{\cmsss Z\kern-.4em Z}} {\lower1.2pt\hbox{\cmsss
Z\kern-.4em Z}}\else{\cmss Z\kern-.4em Z}\fi}

\def\CN {{\cal N}}

\def\CF {{\cal F}}
\def\CJ {{\cal J}}

\def\CL {{\cal L}}

\def\CO {{\cal O}}


\def\CN {{\cal N}}

\def\CO {{\cal O}}

\font\manual=manfnt \def\dbend{\lower3.5pt\hbox{\manual\char127}}

\def\IZ{\relax\ifmmode\mathchoice
{\hbox{\cmss Z\kern-.4em Z}}{\hbox{\cmss Z\kern-.4em Z}}
{\lower.9pt\hbox{\cmsss Z\kern-.4em Z}} {\lower1.2pt\hbox{\cmsss
Z\kern-.4em Z}}\else{\cmss Z\kern-.4em Z}\fi}

\def\bar{\overline}

\def\rt2{\sqrt{2}}
\def\irt2{{1\over\sqrt{2}}}

\def\hat{\widehat}
\def\slashchar#1{\setbox0=\hbox{$#1$}           
  \dimen0=\wd0                                 
  \setbox1=\hbox{/} \dimen1=\wd1               
  \ifdim\dimen0>\dimen1                        
     \rlap{\hbox to \dimen0{\hfil/\hfil}}      
     #1                                        
  \else                                        
     \rlap{\hbox to \dimen1{\hfil$#1$\hfil}}   
     /                                         
  \fi}

\def\foursqr#1#2{{\vcenter{\vbox{
   \hrule height.#2pt
   \hbox{\vrule width.#2pt height#1pt \kern#1pt
   \vrule width.#2pt}
   \hrule height.#2pt
   \hrule height.#2pt
   \hbox{\vrule width.#2pt height#1pt \kern#1pt
   \vrule width.#2pt}
   \hrule height.#2pt
       \hrule height.#2pt
   \hbox{\vrule width.#2pt height#1pt \kern#1pt
   \vrule width.#2pt}
   \hrule height.#2pt
       \hrule height.#2pt
   \hbox{\vrule width.#2pt height#1pt \kern#1pt
   \vrule width.#2pt}
   \hrule height.#2pt}}}}
\def\psqr#1#2{{\vcenter{\vbox{\hrule height.#2pt
   \hbox{\vrule width.#2pt height#1pt \kern#1pt
   \vrule width.#2pt}
   \hrule height.#2pt \hrule height.#2pt
   \hbox{\vrule width.#2pt height#1pt \kern#1pt
   \vrule width.#2pt}
   \hrule height.#2pt}}}}
\def\sqr#1#2{{\vcenter{\vbox{\hrule height.#2pt
   \hbox{\vrule width.#2pt height#1pt \kern#1pt
   \vrule width.#2pt}
   \hrule height.#2pt}}}}

\def\figin{\epsfcheck\figin}\def\figins{\epsfcheck\figins}
\def\epsfcheck{\ifx\epsfbox\UnDeFiNeD
\message{(NO epsf.tex, FIGURES WILL BE IGNORED)}
\gdef\figin##1{\vskip2in}\gdef\figins##1{\hskip.5in}
\else\message{(FIGURES WILL BE INCLUDED)}%
\gdef\figin##1{##1}\gdef\figins##1{##1}\fi}
\def\DefWarn#1{}
\def\figinsert{\goodbreak\midinsert}
\def\ifig#1#2#3{\DefWarn#1\xdef#1{fig.~\the\figno}
\writedef{#1\leftbracket fig.\noexpand~\the\figno}%
\figinsert\figin{\centerline{#3}}\medskip\centerline{\vbox{\baselineskip12pt
\advance\hsize by -1truein\noindent\footnotefont{\bf
Fig.~\the\figno:\ } \it#2}}
\bigskip\endinsert\global\advance\figno by1}

\lref\IntriligatorDD{
  K.~A.~Intriligator, N.~Seiberg and D.~Shih,
  ``Dynamical SUSY breaking in meta-stable vacua,''
  JHEP {\bf 0604}, 021 (2006)
  [arXiv:hep-th/0602239].
}

\lref\SeibergRS{
  N.~Seiberg and E.~Witten,
  ``Monopole Condensation, And Confinement In N=2 Supersymmetric Yang-Mills
  Theory,''
  Nucl.\ Phys.\  B {\bf 426}, 19 (1994)
  [Erratum-ibid.\  B {\bf 430}, 485 (1994)]
  [arXiv:hep-th/9407087].
}

\lref\SeibergAJ{
  N.~Seiberg and E.~Witten,
  ``Monopoles, duality and chiral symmetry breaking in N=2 supersymmetric
  QCD,''
  Nucl.\ Phys.\  B {\bf 431}, 484 (1994)
  [arXiv:hep-th/9408099].
}

\lref\KomargodskiRZ{
  Z.~Komargodski and N.~Seiberg,
  ``From Linear SUSY to Constrained Superfields,''
  JHEP {\bf 0909}, 066 (2009)
  [arXiv:0907.2441 [hep-th]].
}

\lref\FerraraPZ{
  S.~Ferrara and B.~Zumino,
  ``Transformation Properties Of The Supercurrent,''
  Nucl.\ Phys.\  B {\bf 87}, 207 (1975).
}

\lref\KuzenkoAM{
  S.~M.~Kuzenko,
  JHEP {\bf 1004}, 022 (2010)
  [arXiv:1002.4932 [hep-th]].
}

\lref\StelleGI{
  K.~S.~Stelle,
  ``Extended Supercurrents And The Ultraviolet Finiteness Of N=4 Supersymmetric
  Yang-Mills Theory,''
}

\lref\SohniusPK{
  M.~F.~Sohnius,
  ``The Multiplet Of Currents For N=2 Extended Supersymmetry,''
  Phys.\ Lett.\  B {\bf 81}, 8 (1979).
}

\lref\WestTG{
  P.~C.~West,
  ``Introduction to supersymmetry and supergravity,''
{\it  Singapore, Singapore: World Scientific (1990) 425 p}
}

\lref\KuzenkoPI{
  S.~M.~Kuzenko and S.~Theisen,
  ``Correlation functions of conserved currents in N = 2 superconformal
  theory,''
  Class.\ Quant.\ Grav.\  {\bf 17}, 665 (2000)
  [arXiv:hep-th/9907107].
}

\lref\HowePW{
  P.~S.~Howe and P.~C.~West,
  ``Superconformal Ward identities and N = 2 Yang-Mills theory,''
  Nucl.\ Phys.\  B {\bf 486}, 425 (1997)
  [arXiv:hep-th/9607239].
}

\lref\SeibergUR{
  N.~Seiberg,
  ``Supersymmetry And Nonperturbative Beta Functions,''
  Phys.\ Lett.\  B {\bf 206}, 75 (1988).
}

\lref\ArkaniHamedMJ{
  N.~Arkani-Hamed and H.~Murayama,
  ``Holomorphy, rescaling anomalies and exact beta functions in  supersymmetric
  gauge theories,''
  JHEP {\bf 0006}, 030 (2000)
  [arXiv:hep-th/9707133].
}

\lref\AntoniadisVB{
  I.~Antoniadis, H.~Partouche and T.~R.~Taylor,
  ``Spontaneous Breaking of N=2 Global Supersymmetry,''
  Phys.\ Lett.\  B {\bf 372}, 83 (1996)
  [arXiv:hep-th/9512006].
}

\lref\KomargodskiPC{
  Z.~Komargodski and N.~Seiberg,
  ``Comments on the Fayet-Iliopoulos Term in Field Theory and Supergravity,''
  JHEP {\bf 0906}, 007 (2009)
  [arXiv:0904.1159 [hep-th]].
}

\lref\KomargodskiRB{
  Z.~Komargodski and N.~Seiberg,
  ``Comments on Supercurrent Multiplets, Supersymmetric Field Theories and
  Supergravity,''
  arXiv:1002.2228 [hep-th].
}

\lref\HoweTM{
  P.~S.~Howe, K.~S.~Stelle and P.~K.~Townsend,
  ``The Relaxed Hypermultiplet: An Unconstrained N=2 Superfield Theory,''
  Nucl.\ Phys.\  B {\bf 214}, 519 (1983).
}

\lref\FayetYI{
  P.~Fayet,
  ``Fermi-Bose Hypersymmetry,''
  Nucl.\ Phys.\  B {\bf 113}, 135 (1976).
}

\lref\GatesNR{
  S.~J.~Gates, M.~T.~Grisaru, M.~Rocek and W.~Siegel,
  ``Superspace, or one thousand and one lessons in supersymmetry,''
  Front.\ Phys.\  {\bf 58}, 1 (1983)
  [arXiv:hep-th/0108200].
}

\lref\AmbrosettiZA{
  N.~Ambrosetti, I.~Antoniadis, J.~P.~Derendinger and P.~Tziveloglou,
  ``Nonlinear Supersymmetry, Brane-bulk Interactions and Super-Higgs without Gravity,''
  Nucl.\ Phys.\  B {\bf 835}, 75 (2010)
  [arXiv:0911.5212 [hep-th]].
}

\lref\JacotDF{
  J.~C.~Jacot and C.~A.~Scrucca,
  ``Metastable supersymmetry breaking in N=2 non-linear sigma-models,''
  arXiv:1005.2523 [hep-th].
}

\rightline{CERN-PH-TH/2010-103}
\rightline{NSF-KITP-10-058}
\Title{\vbox{\baselineskip12pt }} {\vbox{\centerline{Goldstinos, Supercurrents and Metastable SUSY Breaking} \centerline{in $\CN=2$ Supersymmetric Gauge Theories}}}
\smallskip
\centerline{Ignatios Antoniadis\foot{On leave from CPHT (UMR CNRS 7644) Ecole Polytechnique, F-91128 Palaiseau} and Matthew Buican}
\smallskip
\bigskip
\centerline{{\it Department of Physics, CERN Theory Division, CH-1211 Geneva 23, Switzerland}} %
\vskip 1cm

\noindent We construct an $\CN=2$ supersymmetric generalization of the $\CN=1$ supercurrent formalism of Komargodski and Seiberg (KS) and use it to show that $\CN=2$ theories with linear superconformal anomalies cannot break SUSY under certain broad assumptions. This result suggests that there are no metastable SUSY breaking vacua in a large class of theories that includes $\CN=2$ Super Yang-Mills (SYM).

\bigskip
\Date{May 2010}

\newsec{Introduction}
It is well-known that $\CN=1$ supersymmetric QCD (SQCD) has metastable supersymmetry (SUSY) breaking vacua \IntriligatorDD. However, the situation for $\CN=2$ SQCD is less clear. Indeed, while Seiberg and Witten found an exact description of the quantum moduli space of SUSY vacua in $\CN=2$ SQCD \SeibergRS\SeibergAJ, one expects, in particular, that higher derivative corrections would be important in any (metastable) SUSY breaking vacua. In this paper, we give strong evidence that $\CN=2$ SYM and its cousins do not have metastable SUSY breaking vacua.

Our proof is based on generalizing the results of a recent work by Komargodski and Seiberg \KomargodskiRZ\ (see also \KomargodskiPC\KomargodskiRB). In that paper, the authors showed that the superconformal anomaly multiplet defined in the UV flows, under the renormalization group, to a constrained chiral superfield in the IR that contains the goldstino as a component.

In what follows, we will use the RG evolution of the $\CN=2$ linear superconformal anomaly multiplet to show that as long as an $\CN=2$ QFT has such a multiplet, it cannot contain SUSY breaking vacua that admit a weakly coupled description in terms of goldstinos and goldstone bosons.\foot{It would be interesting to better understand the relation of this statement to the no-go theorems recently discussed in \JacotDF.} We explicitly show that $\CN=2$ SYM falls into this class of theories and therefore conclude that it cannot contain weakly coupled SUSY breaking vacua. In fact, this proof generalizes to arbitrary gauge group and matter content as long as the theory possesses an (at most spontaneously broken) $SU(2)_R$ symmetry.\foot{We would like to emphasize that it is not important whether the theory admits additional superconformal anomaly multiplets. Indeed, as long as the theory has a linear superconformal anomaly multiplet, we are free to study its RG flow. In doing so, we will find the contradictions described below.} We will describe more explicitly some of these extensions after our detailed discussion of $\CN=2$ SYM.

It would also be interesting to extend our analysis to theories that explicitly break $SU(2)_R$. We know that some theories in this class {\it do} admit SUSY breaking vacua since they contain $\CN=2$ D-terms which also explicitly break $SU(2)_R$ (this set of theories includes modifications of the original $\CN=2$ SUSY breaking theory studied by Fayet \FayetYI, as well as models of partial supersymmetry breaking \AntoniadisVB). These theories do not posses linear superconformal anomaly multiplets.

In the next section we briefly review the KS formalism and then quickly proceed to generalize it to the case of interest. We end with some open questions.

\newsec{The KS formalism}
The starting point of the KS formalism is the supercurrent multiplet constructed by Ferrara and Zumino (FZ) \FerraraPZ. In a superconformal theory, the FZ multiplet is described by a real superfield, $\CJ_{\alpha\dot\alpha}$, subject to the constraint
\eqn\FZi{
D^{\alpha}\CJ_{\alpha\dot\alpha}=0.
}
This multiplet contains as its lowest component a conserved $U(1)_R$ current, $j^{\mu}$, while the $\theta$ and $\theta\bar\theta$ components correspond to the supercurrent, $S^{\mu}_{\alpha}$, and the stress-tensor, $T_{\mu\nu}$, respectively. One can describe the breaking of superconformal symmetry in a relatively general way\foot{For various caveats that will not be important to us below, see \KomargodskiRB\ and \KuzenkoAM.} by modifying \FZi\ as follows:
\eqn\FZii{
D^{\alpha}\CJ_{\alpha\dot\alpha}=\bar D_{\dot\alpha}\bar X,
}
where $X$ is a chiral superfield, i.e.
\eqn\Xi{
\bar D_{\dot\alpha}X=0.
}
After solving the equation \FZii\ in components one finds
\eqn\soln{
X=x+\sqrt{2}\theta^{\alpha}\left({\sqrt{2}\over3}\sigma^{\mu}_{\alpha\dot\alpha}\bar S_{\mu}^{\dot\alpha}\right)+\theta^2\left({2\over3}T+i\partial_{\mu}j^{\mu}\right),
}
where the $\theta$ component of $X$ is just the spin-half component of the supercurrent, $T$ is the trace of the stress tensor, and $\partial_{\mu}j^{\mu}$ is the divergence of the $U(1)_R$ current.

The remarkable insight of Komargodski and Seiberg was to realize that in theories of SUSY breaking, the operator $X$, defined in the UV, flows to a well-defined operator, $X_{NL}$, in the deep IR that contains the goldstino. For a generic theory, the scalar partner of the goldstino will acquire a mass. The simplest state such an operator can then create is the two goldstino state. Since supercharge (anti)commutators are well-defined even when SUSY is spontaneously broken, the components of $X$ must satisfy chiral commutation relations along the full RG flow. It is then easy to see that the unique form of $X$, up to an uninteresting constant, is \KomargodskiRZ
\eqn\XNL{
X_{NL}={G^2\over2F}+\sqrt{2}\theta G+\theta^2F,
}
where $G_{\alpha}$ is the goldstino. Additionally, it follows from \XNL\ that $X_{NL}$ satisfies the nilpotent relation
\eqn\Xsq{
X^2_{NL}=0.
}

\newsec{Generalizing the KS formalism to $\CN=2$}
Let us now generalize the above construction to $\CN=2$ SUSY. While there is no unique generalization of the previous discussion to $\CN=2$, we will find that a simple generalization, due originally to Sohnius \SohniusPK\ (see also the more recent discussion in \KuzenkoPI), suffices for our purposes. In particular, we will see that the conservation equation \FZii\ follows from a projection to $\CN=1$ superspace and that the multiplet we consider consists of the two supercurrents, the stress tensor, and the four currents of the $U(2)_R$ $\CN=2$ R-symmetry (along with various other operators required by extended SUSY). Of the four R-currents in the supercurrent multiplet, three of them, corresponding to $SU(2)_R\subset U(2)_R$, will be conserved while the fourth, corresponding to the $\CN=1$ superconformal R-current, will not.\foot{One can also consider a multiplet in which all four $U(2)_R$ currents are conserved \StelleGI. This is a simple generalization of the $\CN=1$ case in which the theory has an exact $U(1)_R$ symmetry.} The reason we choose such a multiplet is simple: it is the precise multiplet structure one finds in $\CN=2$ SYM \StelleGI\WestTG\HowePW.

\subsec{$\CN=2$ SCFT warmup}

We begin our analysis by considering the $\CN=2$ analog of \FZi. It turns out that the supercurrents of an $\CN=2$ superconformal theory can be packaged into a real dimension two superfield, $\CJ$, that satisfies
\eqn\Niisc{
D^{\left<ij\right>}\CJ=\bar D^{\left<ij\right>}\CJ=0,
}
where $i=1,2$ is an index in the fundamental of $SU(2)_R$ and we place $SU(2)_R$ indices inside angular brackets to distinguish them from powers of operators.\foot{In \Niisc\ we define the $SU(2)_R$ spin one differential operator, $D^{\left<ij\right>}$, to be $D^{\left<ij\right>}=D^{\left<ji\right>}\equiv D^{\left<j\right>\alpha}D^{\left<i\right>}_{\alpha}$.} The independent components of $\CJ$ are then
\eqn\tableone{\matrix{& SU(2)_R & {\rm Dim} \cr & \cr  J & {\bf 1} & 2 \cr  J^{\left<i\right>}_{\alpha} & {\bf 2} & 5/2 \cr  J_{\alpha\beta} & {\bf 1} & 3 \cr J^{\left<i\right>}_{\left<j\right>\mu} & {\bf 3} & 3 \cr J_{\mu} & {\bf 1} & 3\cr J^{\left<i\right>}_{\mu\alpha} & {\bf 2} & 7/2 \cr T_{\mu\nu} & {\bf 1} & 4   }}
where we have omitted the hermitian conjugates $\bar J_{\left<i\right>\dot\alpha}$, $\bar J_{\dot\alpha\dot\beta}$, and $\bar J_{\left<i\right>\mu\dot\alpha}$.\foot{In our conventions, $J^{\left<i\right>}_{\left<j\right>\mu}\equiv{1\over2}\sigma^{a \left<i\right>}_{\ \ \left<j\right>}J_{a\mu}$.}

\bigskip
\centerline{\it Reformulation in terms of $\CN=1$ superspace}~
For our discussion below, we will not need the full $\CN=2$ superspace. In fact, it will be convenient to consider the above operators in an explicit $\CN=1$ basis and to define rotations into the remaining half of $\CN=2$ superspace via the action of the second set of supercharges. Therefore, let us consider the following $\CN=1$ projections of $\CJ^{\left<ij\right>}$
\eqn\Niproj{
\hat\CJ\equiv\CJ|, \ \ \ \CJ_{\alpha}\equiv (D^{\left<2\right>}_{\alpha}\CJ)|, \ \ \ \CJ_{\alpha\dot\alpha}=\left(-{1\over3}\left[D^{\left<1\right>}_{\alpha}, \bar D_{\left<1\right>\dot\alpha}\right]+\left[D^{\left<2\right>}_{\alpha}, \bar D_{\left<2\right>\dot\alpha}\right]\right)\CJ|,
}
where the vertical line $\lq\lq |$" denotes the origin of the $(\theta_{\left<2\right>}, \bar\theta^{\left<2\right>})$ superspace. These multiplets obey the following $\CN=1$ constraints descending from the $\CN=2$ constraint in \Niisc
\eqn\consNi{
\bar D^2\hat\CJ=D^2\hat\CJ=0, \ \ \ D^{\alpha}\CJ_{\alpha}=0, \ \ \ \bar D^{2}\CJ_{\alpha}=0, \ \ \ \bar D^{\dot\alpha}\CJ_{\alpha\dot\alpha}=0,
}
where the third constraint follows from the first by applying a variation with respect to the second SUSY. Note that the first equation implies that $\hat\CJ$ is a conserved current multiplet. The second and third equations imply that $\CJ_{\alpha}$ is a conserved current multiplet as well. As we will see momentarily, the conserved currents sitting in the $\theta\bar\theta$ components $\CJ_{\alpha}$ and $\hat\CJ$ are just the second supercurrent and the R-current corresponding to the R-symmetry that leaves the explicit $\CN=1$ superspace invariant.\foot{Since the $\CN=1$ supercurrent does not sit in $\hat\CJ$, the only consistent R-current that can sit in $\hat\CJ$ is the one we have described.}

We can expand the above multiplets in $\CN=1$ superspace as follows
\eqn\ssexp{\eqalign{
&\hat\CJ=J+\theta^{\alpha} J^{\left<1\right>}_{\alpha}+\bar\theta_{\dot\alpha} \bar J_{\left<1\right>}^{\dot\alpha}+\theta\sigma^{\mu}\bar\theta (-{1\over2}J_{\mu}+J^{\left<1\right>}_{\left<1\right>\mu})+\CO(\theta^2\bar\theta, \bar\theta^2\theta)\cr&\CJ_{\alpha}=J^{\left<2\right>}_{\alpha}+\theta^{\beta}J_{\beta\alpha}+\sigma^{\mu}_{\alpha\dot\alpha}\bar\theta^{\dot\alpha} J^{\left<2\right>}_{\left<1\right>\mu}+\theta\sigma^{\mu}\bar\theta (-J_{\mu\alpha}^{\left<2\right>}+{2\over3}i\sigma_{\mu\nu\alpha}^{\ \ \ \beta}\partial^{\nu}J^{\left<2\right>}_{\beta})+\CO(\theta^2\bar\theta, \bar\theta^2\theta)\cr&\CJ_{\mu}={1\over3}J_{\mu}+{4\over3}J^{\left<1\right>}_{\left<1\right>\mu}+\theta^{\alpha}J^{\left<1\right>}_{\mu\alpha}+\bar\theta_{\dot\alpha}J^{\dot\alpha}_{\left<1\right>\mu}+\theta\sigma^{\nu}\bar\theta\left(2T_{\nu\mu}-{1\over4}\epsilon_{\nu\mu\rho\sigma}\partial^{[\rho}j^{\sigma]}\right)+\CO(\theta^2\bar\theta, \bar\theta^2\theta)
}}
where the higher-order terms depend on the lower components.\foot{Note that as in \KomargodskiRZ, the lowest component of $\CJ_{\alpha\dot\alpha}$ is just the superconformal R-current of the manifest $\CN=1$ supersymmetry. Also, in the last equation of \ssexp\ we have defined $\CJ_{\mu}={1\over4}\bar\sigma^{\mu\dot\alpha\alpha}\CJ_{\alpha\dot\alpha}$.}
From the expansions in \ssexp\ and the conservation equations in \consNi, we easily verify the following component conservation equations and trace identities
\eqn\cons{
\partial^{\mu}J^{\left<ij\right>}_{\mu}=0, \ \ \ \partial^{\mu}J_{\mu}=0, \ \ \ \partial^{\mu}J^{\left<i\right>}_{\mu\alpha}=0, \ \ \ \bar\sigma^{\mu\dot\alpha\alpha}J^{\left<i\right>}_{\mu\alpha}=0, \ \ \ \partial^{\mu}T_{\mu\nu}=0, \ \ \ T=0.
}
In fact, one can check that $J^{\left<i\right>}_{\left<j\right>\mu}$ and $J_{\mu}$ are just the conserved superconformal R-currents, $J^{\left<i\right>}_{\mu\alpha}$ are the conserved supercurrents, and $T_{\mu\nu}$ is the conserved and traceless stress-tensor.\foot{As a result, the $\CJ$ multiplet has twenty-four bosonic and twenty-four fermionic components.}

\subsec{The $\CN=2$ linear superconformal anomaly multiplet}
To eventually make contact with SUSY breaking, we must generalize \FZii\ by adding an appropriate representation of $\CN=2$ SUSY to the RHS of \Niisc. For our purposes it suffices to introduce the linear anomaly multiplet originally considered by Sohnius in \SohniusPK\ and so we take
\eqn\Sohnius{
D^{\left<ij\right>}\CJ=3\CL^{\left<ij\right>},
}
with $\CL^{\left<ij\right>}$ being a real $SU(2)_R$ spin one $\CN=2$ linear superfield, i.e. a multiplet satisfying the following conditions
\eqn\defL{
(\CL^{\left<ij\right>})^{\dagger}=\epsilon_{\left<ik\right>}\epsilon_{\left<jl\right>}\CL^{\left<kl\right>}, \ \ \ \CL^{\left<ij\right>}=\CL^{\left<ji\right>}, \ \ \  D_{\alpha}^{(\left<i\right>}\CL^{\left<jk\right>)}=\bar D_{\dot\alpha}^{(\left<i\right>}\CL^{\left<jk\right>)}=0,
}
where $\lq\lq(...)"$ denotes total symmetrization of the included indices. The independent components of $\CL^{\left<ij\right>}$ are then
\eqn\tabletwo{\matrix{& SU(2)_R & {\rm Dim} \cr & \cr  L^{\left<ij\right>} & {\bf 3} & 3 \cr  L^{\left<i\right>}_{\alpha} & {\bf 2} & 7/2 \cr  L_0& {\bf 1} & 4 \cr L_{\mu} & {\bf 1} & 4 }}
Let us stress once more that, as mentioned in the introduction, the theories we study have a conserved (possibly spontaneously broken) $SU(2)_R\subset U(2)_R$ symmetry. Indeed, as we will see below, any solution to \Sohnius\ necessarily satisfies
\eqn\consR{
\partial_{\mu}J^{\left<i\right>\mu}_{\left<j\right>}=0.
}
The intuitive reason for this is that, as we can see from \tabletwo, $\CL^{\left<ij\right>}$ lacks a dimension four $SU(2)_R$ scalar triplet operator.

\bigskip
\centerline{\it The $\CN=2$ linear anomaly multiplet in $\CN=1$ superspace}~
Momentarily, it will prove useful to consider the organization of the fields in $\CL^{\left<ij\right>}$ under the manifest $\CN=1$ SUSY of the previous subsection, and so we define
\eqn\CompCL{
X\equiv\CL^{\left<22\right>}|, \ \ \ L\equiv i\CL^{\left<12\right>}|
}
Notice that the reality condition in \defL\ implies that $X$ is complex while $L$ is real.\foot{The definitions in \CompCL\ and the conservation equation in \defL\ imply that $\left[Q^{\left<2\right>}_{\alpha}, X\right]=0$, $\left[\bar Q_{\left<2\right>\dot\alpha}, X\right]=-2i\bar D_{\dot\alpha}L$, $\left[Q^{\left<2\right>}_{\alpha}, \bar X\right]=2iD_{\alpha}L$, $\left[\bar Q_{\left<2\right>\dot\alpha}, \bar X\right]=0$, $\left[\bar Q_{\left<2\right>\dot\alpha}, L\right]={i\over2}\bar D_{\dot\alpha}\bar X$, and $\left[Q^{\left<2\right>}_{\alpha}, L\right]=-{i\over2}D_{\alpha}X$. 
} Furthermore, the $\CN=2$ current conservation equation in \defL\ descends to the following equations in $\CN=1$ superspace
\eqn\cons{
\bar D_{\dot\alpha} X=0, \ \ \ D^2L=\bar D^2L=0.
}
As a result, we see that $X$ is just a chiral superfield while $L$ is an $\CN=1$ linear superfield. Therefore, the full anomaly multiplet has eight bosonic and eight fermionic components with the following superfield expansion
\eqn\XL{\eqalign{
&X=L^{\left<22\right>}+\theta^{\alpha}L^{\left<2\right>}_{\alpha}+\theta^2L_0+i\theta\sigma^{\mu}\bar\theta\partial_{\mu}L^{\left<22\right>}+\CO(\theta^2\bar\theta)\cr&L=i L^{\left<12\right>}+{i\over2}\theta^{\alpha}L_{\alpha}^{\left<1\right>}-{i\over2}\bar\theta_{\dot\alpha}\bar L^{\dot\alpha}_{\left<1\right>}+i\theta\sigma^{\mu}\bar\theta L_{\mu}+\CO(\theta^2\bar\theta, \bar\theta^2\theta)
}}
where $L^{\mu}$ is the conserved dimension four central charge current.

Analyzing \Sohnius\ in $\CN=1$ superspace, we find the following equations
\eqn\Sohniusi{
\bar D^2\hat\CJ=3X, \ \ \ D^{\alpha}\CJ_{\alpha}=-3iL, \ \ \ \bar D^2\CJ_{\alpha}=0, \ \ \ \bar D^{\dot\alpha}\CJ_{\alpha\dot\alpha}=D_{\alpha}X
}
The third equation in \Sohniusi\ follows from the first equation because $X$ is {\it anti-chiral} with respect to the second SUSY. The fourth equation defines $\CJ_{\alpha\dot\alpha}$ as the FZ multiplet of the explicitly realized $\CN=1$ SUSY.

Solving the second, third, and fourth equations in \Sohniusi\ we find
\eqn\ssexp{\eqalign{
&\CJ_{\alpha}=J^{\left<2\right>}_{\alpha}+\theta^{\beta}\left(J_{\beta\alpha}-{3\over2}i\epsilon_{\beta\alpha}\ell\right)+\sigma^{\mu}_{\alpha\dot\alpha}\bar\theta^{\dot\alpha} J^{\left<2\right>}_{\left<1\right>\mu}+\theta\sigma^{\mu}\bar\theta (-J_{\mu\alpha}^{\left<2\right>}-{1\over2}(\sigma_{\mu}\bar\sigma^{\rho}J^{\left<2\right>}_{\rho})_{\alpha}\cr&+{2\over3}i\sigma_{\mu\nu\alpha}^{\ \ \ \beta}\partial^{\nu}J^{\left<2\right>}_{\beta})+2\theta^2\sigma^{\mu}_{\alpha\dot\alpha}\bar J^{\dot\alpha}_{\left<2\right>\mu}+
\theta^2\bar\theta_{\dot\alpha}\left[{3\over2}\sigma^{\mu\dot\alpha}_{\alpha}\left(-{1\over2}\partial_{\mu}\ell+i\ell_{\mu}\right)+{i\over2}\sigma^{\mu\dot\alpha}_{\beta}\partial_{\mu}J_{\alpha}^{\ \beta}\right]
+\CO(\bar\theta^2\theta)\cr&\CJ_{\mu}=j_{\mu}^{\CN=1}+\theta^{\alpha}\left(J^{\left<1\right>}_{\mu\alpha}+{1\over3}(\sigma_{\mu}\bar\sigma^{\rho}J^{\left<1\right>}_{\rho})_{\alpha}\right)+\bar\theta_{\dot\alpha}\left(\bar J^{\dot\alpha}_{\left<1\right>\mu}+{1\over3}\epsilon^{\dot\alpha\dot\beta}(\bar J_{\left<1\right>\rho}\bar\sigma^{\rho}\sigma_{\mu})_{\dot\beta}\right)\cr&+\theta\sigma^{\nu}\bar\theta\left(2T_{\nu\mu}-{2\over3}\eta_{\mu\nu}T-{1\over4}\epsilon_{\nu\mu\rho\sigma}\partial^{[\rho}j_{\CN=1}^{\sigma]}\right)+{i\over2}\theta^2\partial_{\mu}\bar x-{i\over2}\bar\theta^2\partial_{\mu}x+\CO(\theta^2\bar\theta, \bar\theta^2\theta)
}}
with $X$ and $L$ expressed in terms of these components as follows\foot{In \ssexp\ we have defined $j_{\mu}^{\CN=1}\equiv{1\over3}J_{\mu}+{4\over3}J^{\left<1\right>}_{\left<1\right>\mu}$ and $\ell_{\mu}=iL_{\mu}$.}
\eqn\XLsoln{\eqalign{
&X=x+\theta^{\alpha}\left({2\over3}\sigma^{\mu}_{\alpha\dot\alpha}\bar J_{\left<1\right>\mu}^{\dot\alpha}\right)+\theta^2\left({2\over3}T+{i\over3}\partial_{\mu}J^{\mu}\right)+\CO(\theta\bar\theta)\cr& L=\ell-{i\over2}\theta^{\alpha}\left({2\over3}\sigma^{\mu}_{\alpha\dot\alpha}\bar J_{\left<2\right>\mu}^{\dot\alpha}\right)+{i\over2}\bar\theta_{\dot\alpha}\left({2\over3}\sigma^{\mu\dot\alpha}_{\alpha}J_{\mu}^{\left<2\right>\alpha}\right)+\theta\sigma^{\mu}\bar\theta \ell_{\mu}+\CO(\theta^2\bar\theta, \bar\theta^2\theta).
}}
In writing the $F$ component of $X$, we have used the fact that our solution in \ssexp\ preserves $SU(2)_R$ ($\partial_{\mu}J^{\left<i\right>\mu}_{\left<j\right>}=0$) to rewrite
\eqn\Xcurr{
\partial_{\mu}j^{\mu}_{\CN=1}={1\over3}\partial_{\mu}J^{\mu}.
}
Furthermore, note that, as claimed above, the conserved current in $L$ corresponds to the central charge current since \ssexp\ yields
\eqn\centsch{
\left\{Q_{\alpha}, J^{\left<2\right>\mu}_{\beta}\right\}=3i\epsilon_{\alpha\beta}\ell^{\mu}+\left(2i\sigma^{\mu\nu}_{\alpha\gamma}\partial_{\nu}J_{\beta}^{\gamma}-{2\over3}i\sigma^{\mu\nu}_{\beta\gamma}\partial_{\nu}J^{\gamma}_{\alpha}-2\sigma^{\mu\nu}_{\beta\alpha}\partial_{\nu}\ell\right),
}
where the terms in parenthesis are Schwinger terms.

Of course, the solution we have derived in \ssexp\ is not unique. As stressed in \KomargodskiPC\KomargodskiRZ, there is still the freedom to choose improvement terms. In our case, these improvement terms fill out a reduced $\CN=2$ chiral multiplet, $W$, 
 i.e., a chiral multiplet satisfying the $\CN=2$ generalization of the Bianchi identity,
\eqn\bianchi{
D^{\left<ij\right>}W=\bar D^{\left<ij\right>}\bar W.
}
Indeed, it is easy to check that for any solution $\CJ, \CL^{\left<ij\right>}$ to \Sohnius, the pair
\eqn\shiftpair{
\CJ-{3\over2}(W+\bar W), \ \ \ \CL^{\left<ij\right>}-{1\over2}D^{\left<ij\right>}W,
}
also satisfies the same equations with the component currents shifted by the following improvement terms
\eqn\changes{\eqalign{
&\delta J^{\left<j\right>}_{\left<i\right>\mu}=0, \ \ \ \delta \ell^{\mu}=2\sqrt{2}\partial_{\nu}F^{\mu\nu}, \ \ \ \delta J^{\left<i\right>}_{\mu\alpha}=2i\sigma_{\mu\nu\alpha}^{\ \ \beta}\partial^{\nu}\lambda_{\beta}^{\left<i\right>},\cr& \delta T_{\mu\nu}=-(\partial_{\nu}\partial_{\mu}-\eta_{\mu\nu}\partial^2)(\phi+\bar\phi),
}}
where $\lambda^{\left<i\right>}$ are the two Weyl fermions in $W$, $\phi=W|_{\theta_{\left<i\right>}, \bar\theta^{\left<i\right>}}$ is the lowest component of the multiplet, and $F^{\mu\nu}$ is the anti-symmetric tensor in $W_{\alpha}\equiv\left\{Q^2_{\alpha}, W\right\}|_{\theta_{\left<2\right>},\bar\theta^{\left<2\right>}}$.\foot{The terms given in \changes\ are improvement terms as long as the fields on the RHS vanish sufficiently quickly as we go to infinity.}

\subsec{Spontaneous SUSY breaking}
We would now like to study the solutions to \Sohnius\ as we flow along the RG of a theory that spontaneously breaks the $\CN=2$ SUSY completely. Deep in the IR, the spin half projections of the two supercurrents should flow to the two independent goldstinos
\eqn\SCflow{
S^{\mu \left<i\right>}_{\alpha}=\sqrt{2}f^{\left<i\right>}\sigma^{\mu}_{\alpha\dot\alpha}\bar G^{\left<i\right>\dot\alpha}+...
}
where we can take the $f^{\left<i\right>}$ to be real and positive without loss of generality. Therefore, from \XLsoln\ we see that in the IR $X$ and $L$ contain the two goldstinos as their $\CO(\theta)$ components
\eqn\XLsolnii{\eqalign{
&X_{IR}=x+\sqrt{2}\theta^{\alpha}G_{\left<1\right>\alpha}+\theta^2F+\CO(\theta\bar\theta)\cr& L_{IR}=\ell-{i\over\sqrt{2}}\theta^{\alpha}G_{\left<2\right>\alpha}+{i\over\sqrt{2}}\bar\theta_{\dot\alpha}\bar G^{\left<2\right>\dot\alpha}+\theta\sigma^{\mu}\bar\theta \ell_{\mu}+\CO(\theta^2\bar\theta, \bar\theta^2\theta).
}}
where we have rescaled the superfields by $3/8f$ and redefined the goldstino fields by taking\foot{We give the resulting component SUSY transformations in Appendix A.}
\eqn\Goldstinoredef{
G_{\left<i\right>}\to{f\over f^{\left<i\right>}}G_{\left<i\right>}.
}

It will be useful to remind ourselves again of the $SU(2)_R$ transformation properties of the various components in \XLsolnii. For convenience, we translate table \tabletwo\ into the language of \XLsolnii\ 
\eqn\tablethree{\matrix{& SU(2)_R & {\rm Dim} \cr & \cr  (X, -iL) & {\bf 3} & 1 \cr  G_{\left<i\right>\alpha} & {\bf 2} & 3/2 \cr  F& {\bf 1} & 2 \cr \ell_{\mu} & {\bf 1} & 2 }}
Note that the decay constant sitting in $F$ is an $SU(2)_R$ singlet.

Let us now proceed with our analysis of the IR form of the anomaly multiplet. On general grounds, we expect that unless the bosonic superpartners of the goldstinos in $X_{IR}$ and $L_{IR}$ correspond to goldstone bosons of broken symmetries, they will acquire a mass.

Keeping these facts in mind, first consider the case of unbroken $SU(2)_R$. At zeroth order in an expansion in derivatives, we find that the $\CN=2$ SUSY algebra for the linear multiplet (see Appendix A) fixes the low energy superfields as follows
\eqn\LNLfullzeromom{\eqalign{
&X_{IR}={G_{\left<1\right>}G_{\left<1\right>}\over2F}+{\bar G^{\left<2\right>}\bar G^{\left<2\right>}\over 2\bar F}+\sqrt{2}\theta G_{\left<1\right>}+\theta^2F+\CO(\partial^{\mu}),\cr&L_{IR}=-i{G_{\left<1\right>}G_{\left<2\right>}\over2F}+i{\bar G^{\left<1\right>}\bar G^{\left<2\right>}\over 2\bar F}-{i\over\sqrt{2}}\theta G_{\left<2\right>}+{i\over\sqrt{2}}\bar\theta\bar G^{\left<2\right>}+\CO(\partial^{\mu}).
}}
This solution satisfies the zero momentum (i.e., zero derivative) $\CN=2$ SUSY algebra. However, the explicit solution in \LNLfullzeromom\ does not satisfy the full $\CN=2$ SUSY algebra. Indeed, consider imposing the $\CN=1$ chirality condition, $[\bar\xi \bar Q^{\left<1\right>}, x_{IR}]=0$. We then must demand that at the one derivative level
\eqn\xIRmodified{\eqalign{
&x_{IR}={G_{\left<1\right>}G_{\left<1\right>}\over2F}+{\bar G^{\left<2\right>}\bar G^{\left<2\right>}\over 2\bar F}-{i\over2|F|^2}\left(G_{\left<1\right>}\sigma^{\mu}\bar G^{\left<1\right>}+G_{\left<2\right>}\sigma^{\mu}\bar G^{\left<2\right>}\right)\partial_{\mu}\left({\bar G^{\left<2\right>}\bar G^{\left<2\right>}\over\bar F}\right)\cr&+g_{(0)\dot\alpha}^{\mu}(G_{\left<i\right>}, \bar G^{\left<2\right>})\partial_{\mu}\left({\bar G^{\left<1\right>\dot\alpha}\over\bar F}\right)+g_{(1)}(G_{\left<i\right>}, \bar G^{\left<2\right>})+\CO(\partial^2),
}}
where the terms in $g_{(0)\dot\alpha}^{\mu}$ have zero derivatives and are independent of $\bar G^{\left<1\right>}$. Similarly, the terms in $g_{(1)}$ have one derivative and are independent of $\bar G^{\left<1\right>}$.

On the other hand, imposing the anti-chirality condition with respect to the second SUSY, $\left[\xi Q_{\left<2\right>}, x_{IR}\right]=0$, we find that 
\eqn\xIRmodifiedi{\eqalign{
&x_{IR}={G_{\left<1\right>}G_{\left<1\right>}\over2F}+{\bar G^{\left<2\right>}\bar G^{\left<2\right>}\over 2\bar F}+{i\over2|F|^2}\left(G_{\left<1\right>}\sigma^{\mu}\bar G^{\left<1\right>}+G_{\left<2\right>}\sigma^{\mu}\bar G^{\left<2\right>}\right)\partial_{\mu}\left({G_{\left<1\right>}G_{\left<1\right>}\over F}\right)\cr&+\tilde g_{(0)\mu}^{\alpha}(G_{\left<1\right>}, \bar G^{\left<i\right>})\partial^{\mu}\left({G_{\left<2\right>\alpha}\over F}\right)+\tilde g_{(1)}(G_{\left<1\right>}, \bar G^{\left<i\right>})+\CO(\partial^2),
}}
where the terms in $\tilde g_{(0)\mu}^{\alpha}$ have zero derivatives and are independent of $G_{\left<2\right>}$. Similarly, the terms in $\tilde g_{(1)}$ have one derivative and are independent of $G_{\left<2\right>}$. Clearly, \xIRmodified\ and \xIRmodifiedi\ are not mutually compatible and so we conclude that the goldstinos must have scalar superpartners in the deep IR. Since the SUSY breaking theory was assumed to be interacting, this is unnatural and represents a contradiction.

We should note that in writing \xIRmodified\ and \xIRmodifiedi, we have assumed that the spin one operator in $L_{IR}$ is composite, i.e.
\eqn\lir{
\ell^{\mu}_{IR}=\partial_{\nu}\left({1\over F}G_{\left<1\right>}\sigma^{\mu\nu}G_{\left<2\right>}+{1\over\bar F}\bar 
G^{\left<2\right>}\bar\sigma^{\nu\mu}\bar G^{\left<1\right>}\right)+\CO(\partial^2).
}
One may worry that it is possible to have $\ell^{\mu}$ non-composite in the IR if, for instance, the central charge symmetry is spontaneously broken or if $\ell_{IR}^{\mu}$ is given by the field strength of an abelian gauge field of an unbroken gauge symmetry. While this situation is certainly possible, it does not affect our above reasoning. Indeed, we would find that \xIRmodified\ and \xIRmodifiedi\ are essentially the same up to terms proportional to the difference of the right and left hand sides of \lir, and so our logic is not affected.

It follows from the above analysis that if the goldstinos of $\CN=2$ spontaneous SUSY breaking belong to a linear superfield of the type $\CL^{\left<ij\right>}$, or equivalently $(L,X)$, there are no constraints compatible with $\CN=2$ SUSY and $SU(2)_R$ that can eliminate all remaining components. In principle, this should be possible to show directly using $\CN=2$ superfields which we do not use here.\foot{A possible starting point would be to use the off-shell formulation of the $\CN=2$ linear superfield used in \AmbrosettiZA\ which admits also a constraint with one non-linear SUSY.} The next step would be to weaken our requirement on the IR theory, by including other massless degrees of freedom besides the goldstinos and $\ell^{\mu}_{IR}$. However, as we show below, in that case, although weaker constraints might be compatible with $\CN=2$ SUSY, the symmetries of the low energy theory turn out to be different from those in the UV.

For instance, the previous reasoning does not necessarily go through when the $SU(2)_R$ symmetry is spontaneously broken. Indeed, in such a case, the lowest components of $X_{IR}$ and $L_{IR}$ can contain the R-axions of the broken $SU(2)_R$ since their dimensions and quantum numbers are the same.

However, it is still useful to ask whether we can construct a consistent low energy effective action using $X_{IR}$ and $L_{IR}$ (with the R-axions now in the lowest components). In order to make this question meaningful, we will assume that whatever strong dynamics takes place in the theory can be integrated out and what is left in the deep IR is a weakly coupled theory of goldstinos and Goldstone bosons. In such a case, we expect
\eqn\ActionIR{
\CL_{IR}=\int d^4\theta\left(X_{IR}\bar X_{IR}-2L_{IR}^2\right)+\left(\int d^2\theta fX_{IR}+c.c.\right)+...
}
where the ellipses contain weak interactions and kinetic terms for the other light fields.

In order for \ActionIR\ to constitute a consistent low energy description of our original UV theory, it must be the case that the superconformal anomaly multiplet for \ActionIR\ matches the (RG evolved) superconformal anomaly multiplet for the UV theory. In other words, the anomaly multiplet for \ActionIR\ should just be the pair $(X_{IR}, L_{IR})$ up to possible field redefinitions. However, this is not what happens. The heuristic reason for this difference is that the action for the linear multiplet has a different set of symmetries than the original theory. In particular, there is no (non-trivial) conserved central charge current.

To make our discussion more precise, consider the $\CN=2$ supercurrent for the theory in \ActionIR 
\eqn\SCNii{
\CJ=2\left(\CL^{\left<11\right>}\CL^{\left<22\right>}-\CL^{\left<12\right>}\CL^{\left<12\right>}\right).
}
From this expression, we then deduce the following component conservation equations 
\eqn\SCNiicomp{\eqalign{
&\bar D^2\hat\CJ=8fX_{IR}+2\bar D^2(L_{IR}^2),\cr&D^{\alpha}\CJ_{\alpha}=2iD^{\alpha}L_{IR}D_{\alpha}X_{IR}-8ifL_{IR},\cr&\bar D^2\CJ_{\alpha}=-2i\bar D^2\left(L_{IR}D_{\alpha} X_{IR}\right),\cr&\bar D^{\dot\alpha}\CJ_{\alpha\dot\alpha}={8\over3}D_{\alpha}\left(fX_{IR}-\bar D_{\dot\alpha}L_{IR}\bar D^{\dot\alpha}L_{IR}\right)+2\bar D^2 D_{\alpha}\left(L_{IR}^2\right)+2\bar D_{\dot\alpha}D_{\alpha}X_{IR}\bar D^{\dot\alpha}\bar X_{IR}.
}}
Notice that this set of equations cannot be brought to the form given in \Sohniusi\ with $(X, L)\to (X_{IR}, L_{IR})$. Indeed, suppose we try to shift $\hat\CJ$ so that it satisfies the first conservation equation in \Sohniusi. After performing this transformation, we would still find that $D^{\alpha}\CJ_{\alpha}$ is not a real linear superfield. This mismatch is a symptom of the fact that the symmetry structure of the IR theory is not compatible with the symmetry structure of the original theory.

\newsec{Metastable vacua in $\CN=2$ SYM}
In the previous section, we saw that any $\CN=2$ theory with a linear superconformal anomaly cannot break SUSY under the assumption of  a weakly coupled description in the deep IR.
As we will now see, one interesting consequence of the above proof is that if this assumption holds for $\CN=2$ SYM, it cannot contain metastable SUSY breaking vacua. 

To understand this statement, we begin by noting that in the asymptotically free regime, dimensional analysis and $U(2)_R$ covariance tell us that, up to an overall constant, the SYM supercurrent conservation equation should be (see also the discussions in \HowePW\StelleGI\WestTG)
\eqn\SCanomYM{
D^{\left<ij\right>}\CJ=\CL^{\left<ij\right>}_{{\bf C}, {\rm SYM}}={c\over2}\bar D^{\left<ij\right>}{\rm tr}\bar W^2
}
where $c=8\pi i\beta$ is the one-loop beta function \SeibergUR\ and $W$ is the SYM field strength superfield. Notice that the anomaly satisfies
\eqn\SCanomYMi{
D^{(\left<i\right>}_{\alpha}\CL^{\left<jk\right>)}_{{\bf C}{\rm SYM}}=\bar D^{(\left<i\right>}_{\dot\alpha}\CL^{\left<jk\right>)}_{{\bf C}{\rm SYM}}=0
}
and so it is an $\CN=2$ linear superfield. However, unlike the linear superfield described in the previous section, this superfield is complex (hence the subscript \lq\lq${\bf C}$") and so
\eqn\SCanomYMii{
(\CL^{\left<ij\right>}_{{\bf C}{\rm SYM}})^{\dagger}\ne\epsilon_{\left<ik\right>}\epsilon_{\left<jl\right>}\CL^{\left<kl\right>}_{{\bf C} {\rm SYM}}
}
In particular, \SCanomYM\ and \SCanomYMii\ imply that $\CJ$ embeds all of the conserved currents considered in the previous section plus an additional central charge current (i.e., the central charge current is now complex).

However, as mentioned by the authors in \KuzenkoPI, \SCanomYM\ can be brought to the form \Sohnius\ by the local gauge invariant shift $\CJ\to\CJ-{c\over2}{\rm tr}W^2+{c\over2}{\rm tr}\bar W^2$.
In particular, we find
\eqn\SCanomYMfinal{
D^{\left<ij\right>}\CJ={c\over2}(\bar D^{\left<ij\right>}{\rm tr}\bar W^2-D^{\left<ij\right>}{\rm tr}W^2)
}
Therefore, it follows from the results of the previous section that $\CN=2$ SYM has no metastable vacua as long as the theory has a  weakly coupled description in the IR. Alternatively, one can use arguments analogous to those given in the previous section for the complex anomaly multiplet in \SCanomYM\ and reach the same conclusions.

\subsec{Beyond $\CN=2$ SYM}
We can now conjecture a generalization of this argument. Indeed, notice that the particular $\CN=2$ representation appearing on the RHS of \SCanomYMfinal\ is fixed by the symmetries of the theory and the requirement that the equation be well-defined. Therefore, it seems plausible that a whole host of theories like $\CN=2$ SQCD (and various quiver generalizations) with arbitrary gauge group and number of flavors also cannot have weakly-coupled metastable vacua since these theories have supercurrent multiplets with the same conserved symmetries (conserved $SU(2)_R$ and central charge currents) as $\CN=2$ SYM.

More precisely, consider defining a $\CJ$ as in \SCanomYM\ for an arbitrary non-abelian gauge group (possibly a product gauge group consisting of many non-abelian factors) and adding matter hypermultiplets $Q_i, \tilde Q^i$ transforming in an arbitrary representation of the gauge group so that the theory is still asymptotically free. Let us also assume that we do not deform the theory in the UV by introducing explicit $SU(2)_R$ breaking (note that we cannot include an FI term). Then, it is easy to see that $\CJ$ in \SCanomYM\ should be deformed as follows
\eqn\SCanomSQCD{
\CJ\to\CJ+{1\over2}\sum_i\left(Q_i\bar Q^i+\bar{\tilde{Q_i}}\tilde Q^i\right).
}
Therefore, the theory admits a linear superconformal anomaly and cannot posses weakly coupled SUSY breaking vacua.

\newsec{Conclusions}
We have shown that theories with an $\CN=2$ linear superconformal anomaly cannot break SUSY as long as they admit a free description in the IR. This fact strongly suggests that a large class of theories that includes $\CN=2$ SYM cannot have metastable SUSY breaking vacua.

An interesting future direction would be to study generalizations of the ansatz in \Sohnius\ to include anomalies that describe the breaking of more symmetries (like the $SU(2)_R$ symmetries). For example, one could imagine considering theories in which 
\eqn\conseq{
D^{\left<ij\right>}\CJ=3\CL^{\left<ij\right>},
}
with real but non-conserved $\CL^{ij}$, i.e.
\eqn\consLIJ{
D^{(\left<i\right>}_{\alpha}\CL^{\left<jk\right>)}=D_{\alpha \left<l\right>}\CT^{\left<lijk\right>}, \ \ \ D^{(\left<i\right>}_{\alpha}\CT^{\left<jklm\right>)}=\bar D^{(\left<i\right>}_{\dot\alpha}\bar\CT^{\left<jklm\right>)}=0
}
where $\CT^{\left<ijkl\right>}$ is a real $SU(2)_R$ spin two field.\foot{The pair $(\CL^{\left<ij\right>}, \CT^{\left<ijlm\right>})$ define the so-called \lq\lq relaxed hypermultiplet" \HoweTM.}

\bigskip
\bigskip\centerline{\bf Acknowledgements}
We are grateful to Z. Komargodski and N. Seiberg for interesting comments and discussions. This work was supported in part by the European Commission under the ERC Advanced Grant 226371 and the contract PITN-GA-2009-237920. I.A. was also supported in part by the CNRS grant GRC APIC PICS 3747 and in part by the National Science Foundation under Grant No. PHY05-51164; he would like to thank the KITP of UC Santa Barbara for hospitality during the last part of this work.

\vfill\eject
\appendix{A}{Component field commutators}
The variations of the component fields in the $\CN=2$ linear anomaly multiplet, $\CL_{IR}^{ij}$, with respect to the first SUSY are
\eqn\commutators{\eqalign{
&\left[\xi Q^{\left<1\right>}, x_{IR}\right]=\sqrt{2}\xi G_{\left<1\right>}, \ \ \ \left[\bar\xi\bar Q_{\left<1\right>}, x_{IR}\right]=0\cr&\left[\xi Q^{\left<1\right>}, G_{{\left<1\right>}\alpha}\right]=\sqrt{2}\xi_{\alpha}F, \ \ \ \left[\bar\xi\bar Q_{\left<1\right>}, G_{{\left<1\right>}\alpha}\right]=i\sqrt{2}\sigma^{\mu}_{\alpha\dot\alpha}\bar\xi^{\dot\alpha}\partial_{\mu}x_{IR}\cr&\left[\xi Q^{\left<1\right>}, F\right]=0, \ \ \ \left[\bar\xi \bar Q_{\left<1\right>}, F\right]=i\sqrt{2}\bar\xi_{\dot\alpha}\bar\sigma^{\mu\dot\alpha\alpha}\partial_{\mu}G_{{\left<1\right>}\alpha}\cr&\left[\xi Q^{\left<1\right>}, \bar x_{IR}\right]=0, \ \ \ \left[\bar\xi\bar Q_{\left<1\right>}, \bar x_{IR}\right]=\sqrt{2}\bar\xi\bar G^{\left<1\right>}\cr&\left[\xi Q^{\left<1\right>}, \bar G^{\left<1\right>}_{\dot\alpha}\right]=-i\sqrt{2}\xi^{\alpha}\sigma^{\mu}_{\alpha\dot\alpha}\partial_{\mu}\bar x_{IR}, \ \ \ \left[\bar\xi\bar Q_{\left<1\right>}, \bar G_{\dot\alpha}^{\left<1\right>}\right]=\sqrt{2}\bar\xi_{\dot\alpha}\bar F\cr& \left[\xi Q^{\left<1\right>}, \bar F\right]=i\sqrt{2}\xi^{\alpha}\sigma^{\mu}_{\alpha\dot\alpha}\partial_{\mu}\bar G^{{\left<1\right>}\dot\alpha}, \ \ \ \left[\bar\xi\bar Q_{\left<1\right>}, \bar F\right]=0\cr&\left[\xi Q^{\left<1\right>}, \ell_{IR}\right]=-{i\over\sqrt{2}}\xi G_{{\left<2\right>}}, \ \ \ \left[\bar\xi\bar Q_{\left<1\right>}, \ell_{IR}\right]={i\over\sqrt{2}}\bar\xi\bar G^{{\left<2\right>}}\cr&\left[\xi Q^{\left<1\right>}, G_{{\left<2\right>}\alpha}\right]=0, \ \ \ \left[\bar\xi\bar Q_{\left<1\right>}, G_{{\left<2\right>}\alpha}\right]=i\sqrt{2}\sigma^{\mu}_{\alpha\dot\alpha}\bar\xi^{\dot\alpha}(\ell_{IR \mu}+i\partial_{\mu}\ell_{IR})\cr&\left[\xi Q^{\left<1\right>}, \bar G^{\left<2\right>}_{\dot\alpha}\right]=-i\sqrt{2}\xi^{\alpha}\sigma^{\mu}_{\alpha\dot\alpha}(\ell_{IR \mu}-i\partial_{\mu}\ell_{IR}), \ \ \ \left[\bar\xi \bar Q_{\left<1\right>}, \bar G^{\left<2\right>}_{\dot\alpha}\right]=0\cr&\left[\xi Q^{\left<1\right>}, \ell_{IR}^{\mu}\right]=\sqrt{2}\xi\sigma^{\mu\nu}\partial_{\nu}G_{\left<2\right>},  \ \ \ \left[\bar\xi\bar Q_{\left<1\right>}, \ell_{IR}^{\mu}\right]=-\sqrt{2}\partial_{\nu}\bar G^{\left<2\right>}\bar\sigma^{\mu\nu}\bar\xi
}}
The variations of the component fields with respect to the second SUSY are
\eqn\commutatorsii{\eqalign{
&\left[\xi Q^{\left<2\right>}, x_{IR}\right]=0, \ \ \ \left[\bar\xi\bar Q_{\left<2\right>}, x_{IR}\right]=\sqrt{2}\bar\xi\bar G^{\left<2\right>}\cr&\left[\xi Q^{\left<2\right>}, G_{{\left<1\right>}\alpha}\right]=0, \ \ \ \left[\bar\xi\bar Q_{\left<2\right>}, G_{{\left<1\right>}\alpha}\right]=-i\sqrt{2}\sigma^{\mu}_{\alpha\dot\alpha}\bar\xi^{\dot\alpha}(\ell_{IR\mu}-i\partial_{\mu}\ell_{IR})\cr&\left[\xi Q^{\left<2\right>}, F\right]=0, \ \ \ \left[\bar\xi \bar Q_{\left<2\right>}, F\right]=i\sqrt{2}\bar\xi\bar\sigma^{\mu}\partial_{\mu}G_{\left<2\right>}\cr&\left[\xi Q^{\left<2\right>}, \bar x_{IR}\right]=\sqrt{2}\xi G_{\left<2\right>}, \ \ \ \left[\bar\xi\bar Q_{\left<2\right>}, \bar x_{IR}\right]=0\cr&\left[\xi Q^{\left<2\right>}, \bar G^{\left<1\right>}_{\dot\alpha}\right]=i\sqrt{2}\xi^{\alpha}\sigma^{\mu}_{\alpha\dot\alpha}(\ell_{IR \mu}+i\partial_{\mu}\ell_{IR}), \ \ \ \left[\bar\xi\bar Q_{\left<2\right>}, \bar G_{\dot\alpha}^{\left<1\right>}\right]=0\cr& \left[\xi Q^{\left<2\right>}, \bar F\right]=i\sqrt{2}\xi\sigma^{\mu}\partial_{\mu}\bar G_{\left<2\right>} , \ \ \ \left[\bar\xi\bar Q_{\left<2\right>}, \bar F\right]=0\cr&\left[\xi Q^{\left<2\right>}, \ell_{IR}\right]=-{i\over\sqrt{2}}\xi G_{\left<1\right>}, \ \ \ \left[\bar\xi\bar Q_{\left<2\right>}, \ell_{IR}\right]={i\over\sqrt{2}}\bar\xi\bar G^{\left<1\right>}\cr&\left[\xi Q^{\left<2\right>}, G_{{\left<2\right>}\alpha}\right]=\sqrt{2}\xi_{\alpha}F, \ \ \ \left[\bar\xi\bar Q_{\left<2\right>}, G_{{\left<2\right>}\alpha}\right]=i\sqrt{2}\sigma^{\mu}_{\alpha\dot\alpha}\bar\xi^{\dot\alpha}\partial_{\mu}\bar x_{IR}\cr&\left[\xi Q^{\left<2\right>}, \bar G^{\left<2\right>}_{\dot\alpha}\right]=-i\sqrt{2}\xi^{\alpha}\sigma^{\mu}_{\alpha\dot\alpha}\partial_{\mu}x_{IR}, \ \ \ \left[\bar\xi \bar Q_{\left<2\right>}, \bar G^{\left<2\right>}_{\dot\alpha}\right]=\sqrt{2}\bar\xi_{\dot\alpha}\bar F\cr&\left[\xi Q^{\left<2\right>}, \ell_{IR}^{\mu}\right]=-\sqrt{2}\xi\sigma^{\mu\nu}\partial_{\nu}G_{\left<1\right>},  \ \ \ \left[\bar\xi\bar Q_{\left<2\right>}, \ell_{IR}^{\mu}\right]=\sqrt{2}\partial_{\nu}\bar G^{\left<1\right>}\bar\sigma^{\mu\nu}\bar\xi
}}

At zero momentum (i.e., no derivatives), the algebra simplifies considerably, and the only non-vanishing commutators are 
\eqn\commutatorszeromomi{\eqalign{
&\left[\xi Q^{\left<1\right>}, x_{IR}\right]=\sqrt{2}\xi G_{\left<1\right>}, \ \ \ \left[\xi Q^{\left<1\right>}, G_{{\left<1\right>}\alpha}\right]=\sqrt{2}\xi_{\alpha}F, \ \ \ \left[\bar\xi\bar Q_{\left<1\right>}, \bar x_{IR}\right]=\sqrt{2}\bar\xi\bar G^{\left<1\right>}\cr&\left[\bar\xi\bar Q_{\left<1\right>}, \bar G_{\dot\alpha}^{\left<1\right>}\right]=\sqrt{2}\bar\xi_{\dot\alpha}\bar F, \ \ \ \left[\xi Q^{\left<1\right>}, \ell_{IR}\right]=-{i\over\sqrt{2}}\xi G_{{\left<2\right>}}, \ \ \  \left[\bar\xi\bar Q_{\left<1\right>}, \ell_{IR}\right]={i\over\sqrt{2}}\bar\xi\bar G^{{\left<2\right>}}
}}
and
\eqn\commutatorszeromomii{\eqalign{
&\left[\bar\xi\bar Q_{\left<2\right>}, x_{IR}\right]=\sqrt{2}\bar\xi\bar G^{\left<2\right>}, \ \ \ \left[\xi Q^{\left<2\right>}, \bar x_{IR}\right]=\sqrt{2}\xi G_{\left<2\right>}, \ \ \ \left[\xi Q^{\left<2\right>}, \ell_{IR}\right]=-{i\over\sqrt{2}}\xi G_{\left<1\right>}\cr& \left[\bar\xi\bar Q_{\left<2\right>}, \ell_{IR}\right]={i\over\sqrt{2}}\bar\xi\bar G^{\left<1\right>}, \ \ \ \left[\xi Q^{\left<2\right>}, G_{{\left<2\right>}\alpha}\right]=\sqrt{2}\xi_{\alpha}F, \ \ \ \left[\bar\xi \bar Q_{\left<2\right>}, \bar G^{\left<2\right>}_{\dot\alpha}\right]=\sqrt{2}\bar\xi_{\dot\alpha}\bar F
}}

\appendix{B}{Some simple examples}~
In this appendix we will consider some simple examples to illustrate the above discussion. To that end, consider the following $\CN=2$ $U(1)$ gauge theory with $\CN=2$ field strength $W=(\Phi, W_{\alpha})$
\eqn\Niiqed{
\CL=\int d^2\theta_1d^2\theta_2\ \CF(W)+h.c.
}
where
\eqn\CFdefn{
\CF(W)={1\over4}W^2+...
}
is the $\CN=2$ prepotential and the ellipses contain possible higher-order interactions. Clearly, if the higher-order interactions are non-vanishing, the theory has a superconformal anomaly.

We know that this example must fall within the class of theories we have considered because it appears, for example, in the low energy theory on the Coulomb branch of $SU(2)$ SYM \SeibergRS.\foot{Note also that the expression in \Niiqed\ makes it clear that the theory is $SU(2)_R$ invariant for arbitrary $\CF$. Indeed, $\CF$ is itself $SU(2)_R$ invariant and so too is the $\CN=2$ chiral integration measure.} To see this fact more explicitly, we can construct the $\CN=2$ supercurrent multiplet and anomaly by starting with the operator
\eqn\hatJqed{
\hat J=-\Phi\partial_{\bar\Phi}\bar\CF-\bar\Phi\partial_{\Phi}\CF-\left(\tilde\CF+\bar{\tilde\CF}\right).
}
The first two terms in \hatJqed\ are (minus) the K\"ahler potential for $\Phi$ that descends from \Niiqed,\CFdefn, and the two terms in the parenthesis are local chiral and antichiral shifts. We define the chiral function $\tilde\CF$, up to an uninteresting constant, as follows
\eqn\tildeCFdefn{
\partial_{\Phi}\tilde\CF\equiv\partial_{\Phi}\CF-\Phi\partial^2_{\Phi}\CF.
}
From \hatJqed\ we construct the full supercurrent multiplet by the action of the second set of supercharges. The crucial point is that the anomaly multiplet is just
\eqn\anommult{\eqalign{
&X=-{1\over3}\left(\left(\partial_{\Phi}\CF-\Phi\partial^2_{\Phi}\CF\right)\bar D^2\bar\Phi+2\Phi\partial^3_{\Phi}\CF \ W^2+\bar D^2\bar{\tilde\CF}\right)\cr&L={\sqrt{2}\over3}\left(\bar D_{\dot\alpha}\left[\bar W^{\dot\alpha}\left(\partial_{\bar\Phi}\bar\CF-\bar\Phi\partial^2_{\bar\Phi}\bar\CF\right)\right]+D^{\alpha}\left[W_{\alpha}\left(\partial_{\Phi}\CF-\Phi\partial^2_{\Phi}\CF\right)\right]\right)
}}
Clearly $X$ is chiral and $L$ is real and linear. Therefore, this example is in our class of theories, and it is easy to see  directly that SUSY cannot be broken in this case. Finally, notice that for a trivial prepotential the theory is superconformal and $X=L=0$.
\bigskip
\centerline{\it Massive hypermultiplets}~
For our next example, let us consider a simple generalization of the original example in \SohniusPK\ and examine a theory of free massive hypermultiplets
\eqn\freeMH{
\CL=\int d^4\theta\left(\bar Q^iQ_i+\tilde Q^i\bar{\tilde Q}_i\right)+\left(\int d^2\theta {1\over\sqrt{2}}M_iQ_i\tilde Q^i+h.c.\right).
}
Without loss of generality we can take the masses to be real.
The hypermultiplets are doublets under $SU(2)_R$ and have charge $-1$ under the R-symmetry that leaves the manifest $\CN=1$ superspace invariant. Therefore, we can construct the following $\hat J$ operator for the theory
\eqn\Jhatophyper{
\hat J={1\over2}\sum_i\left(Q_i\bar Q^i+\bar{\tilde Q}_i\tilde Q^i\right).
}
From \Jhatophyper\ and the action of the second SUSY it follows that the anomaly multiplet is
\eqn\hyperanom{\eqalign{
&X={2\sqrt{2}\over3}\sum_iM_iQ_i\tilde Q^i,\cr& L=-{\sqrt{2}\over3}\sum_iM_i\left(Q_i\bar Q^i-\bar{\tilde Q}_i\tilde Q^i\right).
}}
This theory is therefore in the class we have considered in this paper, and it is straight forward to see directly that SUSY is not broken. As a final note, we see that in the limit of vanishing masses, the theory is superconformal and $X=L=0$.

We can also consider coupling the massive hypermultiplets to an $\CN=2$ $U(1)$ gauge multiplet
\eqn\freeMHqed{\eqalign{
\CL&=\int d^4\theta\left(\bar\Phi\Phi+\sum_i\left(\bar Q^ie^{n_iV}Q_i+\tilde Q^ie^{-n_iV}\bar{\tilde Q}_i\right)\right)\cr&+\left(\int d^2\theta\left({1\over4}W^2+{n_i\over\sqrt{2}}\left(\Phi+M_i\right)Q_i\tilde Q^i\right)+h.c.\right),
}}
where we have taken the prepotential to be trivial for simplicity. Using the fact that the free superconformal R-charge of $\Phi$ under the $U(1)_R$ symmetry that leaves the $\CN=1$ superspace invariant is $+2$ we can construct the $\hat J$ operator
\eqn\Jhatophyperi{
\hat J=-\bar\Phi\Phi+{1\over2}\sum_i\left(Q_i\bar Q^i+\bar{\tilde Q}_i\tilde Q^i\right).
}
It then follows that the (classical) anomaly multiplet is just
\eqn\hyperanomi{\eqalign{
&X_{cl}={2\sqrt{2}\over3}\sum_{i}(M_i-M_1)Q_i\tilde Q^i,\cr& L_{cl}=-{\sqrt{2}\over3}\sum_{i}(M_i-M_1)\left(Q_i\bar Q^i-\bar{\tilde Q}_i\tilde Q^i\right).
}}
Without loss of generality, we have shifted the definition of $\Phi$ to $\Phi+M_1$ so as to absorb $M_1\ne0$ into the vacuum expectation value (VEV) of $\Phi$. By examining \Jhatophyperi\ we see that this corresponds to an improvement transformation since we are only considering a trivial prepotential for $\Phi$ (in the quantum theory, with a non-trivial prepotential, this is no longer true, but we consider the classical case for simplicity). Since the VEV of $\Phi$ breaks the superconformal symmetry spontaneously, it does not contribute to the classical anomaly.

This example is also clearly in our class of theories, and it is easy to check directly that SUSY is not broken.

\listrefs
\end